\documentclass[twocolumn,showpacs,preprintnumbers,amsmath,amssymb,showlabels]{revtex4}

\usepackage{graphicx}
\usepackage{dcolumn}
\usepackage{bm}

\begin{document}

\title{Wetting, Spreading, and Adsorption on Randomly Rough Surfaces}

\author{S. Herminghaus}

\affiliation{Max-Planck-Institute for Dynamics and Self-Organization, Bunsenstr. 10, 37073 G\"ottingen, Germany}

\date{\today}

\begin{abstract}
The wetting properties of solid substrates with customary (i.e., macroscopic) random roughness are considered as a function of the microscopic contact angle of the wetting liquid and its partial pressure in the surrounding gas phase. Analytic expressions are derived which allow for any given lateral correlation function and height distribution of the roughness to calculate the wetting phase diagram, the adsorption isotherms, and to locate the percolation transition in the adsorbed liquid film. Most features turn out to depend only on a few key parameters of the roughness, which can be clearly identified. It is shown that a first order transition in the adsorbed film thickness, which we term 'Wenzel prewetting', occurs generically on typical roughness topographies, but is absent on purely Gaussian roughness. It is thereby shown that even subtle deviations from Gaussian roughness characteristics may be essential for correctly predicting even qualitative aspects of wetting. 
\end{abstract}


\maketitle

\section{Introduction} 

While the physics of wetting and spreading on ideally smooth solid surfaces has meanwhile reached a status of mature textbook knowledge, a whole range of wetting phenomena on randomly rough substrates are still elusive. This is particularly annoying, as almost all surfaces of practical interest bear considerable roughness, be it due to weathering, wear, or on purpose as, e.g., in the case of sand-blasted surfaces. Clearly, this has substantial impact in many situations. For example, a drop of  liquid deposited on a rough substrate will spread or not, depending on the morphology of the liquid film which develops in the troughs of the roughness. As it accommodates its free surface to the substrate topography, it may  percolate across the sample. The drop will then gradually spread over the entire sample. On the contrary, if the film rather tends to form isolated domains, the drop will stay in place. Similarly, the redistribution of liquid within a granular pile, such as in humid soil or sand, may proceed along the grain surfaces only if the liquid wetting film on the grains forms a percolated structure. The morphology of a liquid water film deposited from humid air onto the surface of an electric isolator will strongly affect the performance of the latter, for analogous reasons. 
   
There has been already a lot of research on the adsorption of liquids on rough surfaces  
[1-15], but this was concerned either with roughness amplitudes as small as the (nanometer) range of van der Waals forces  \cite{AndelRob1988,KardInd1990,Netz1997,Seemann2001} or with rather artificial substrate topographies in the context of super-hydrophobicity \cite{Herminghaus2000,BicoQuere2001,BicoThiele2002,OkuQuere2004,He2011}, or with completely wetting liquids (zero contact angle) \cite{Philip1978,ColeKrim1989,PalasKrim1993,Seemann2001}. The most frequently encountered, customary case, however, is characterized by a finite contact angle, and a random roughness with typical length scales at least in the micron range. In the present paper, we consider surfaces which exhibit a random topography on scales large as compared to molecules, and are subject to adsorption of a liquid which forms a small but finite contact angle with the substrate material.  Since we consider macroscopic roughness, we adopt the view that all interfaces are infinitely sharp on the length scale of consideration (sharp kink approximation \cite{DietrichReview}). As the typical length scales considered here are still small as compared to the capillary length of the liquid (2.7 mm for water), gravity will be neglected as regards its effect on the liquid surface morphologies to be described. As opposed to earlier studies which concentrated on the macroscopic contact angle and contact line \cite{Joanny1984,Chow1998}, we will try to derive the wetting phase diagram and other characteristics connected to the adsorption of a liquid film. 

 The first systematic study of wetting on a randomly rough substrate at finite contact angle owes to Wenzel \cite{Wenzel1936}. He characterized the roughness by a single parameter, ${\rm r}$, which he defined as the ratio of the total substrate area divided by the projected area. Obviously, ${\rm r} \ge 1$, and  ${\rm r}=1$ corresponds to a perfectly smooth surface. The free energy which is gained per unit area when the rough substrate is covered with a liquid is then given by ${\rm r} (\gamma_{sg}-\gamma_{sl})$, where $\gamma_{sl}$ and $\gamma_{sg}$ are the solid-liquid and solid-gas interfacial tension, respectively. If this is larger than the surface tension of the liquid, $\gamma$, we expect a vanishing macroscopic contact angle, because covering the substrate with liquid releases more energy than is required for the formation of a free liquid surface of the same (projected) area. More specifically, force balance at the three-phase contact line yields
\begin{equation}
\cos\theta_{macro} = \frac{{\rm r} (\gamma_{sg}-\gamma_{sl})}{\gamma} = {\rm r} \cos\theta
\label{Eq:Wenzel}
\end{equation}
for the macroscopic contact angle on the rough surface. $\theta$ is the microscopic contact angle according to Young and Dupr\'e. When the microscopic contact angle is reduced to $\theta_W = \arccos (1/{\rm r})$, which we will henceforth call Wenzel's angle, $\theta_{macro}$ vanishes, and the substrate is covered with an 'infinitely' thick liquid film. As we will see below, however, there are imprortant ramifications which are sensitive to the kind of roughness of the substrate. Furthermore, even minute deviations from liquid-vapour coexistence, as they are omnipresent in practical situations, unveil a rather complex scenario which goes well beyond eq.~(\ref{Eq:Wenzel}).  
 
\section{Presentation of the problem}

We describe the topography of the rough solid substrate by $f(x,y)$, where $(x,y)$ is a vector in the plane. The (randomly varying) function $f$ is normalized such that $<f> = 0$, where the angular brackets denote averaging over the entire sample area, $\mathcal{S}$. It is assumed that the substrate is homogeneous and isotropic, in the sense that the statistical parameters of $f$ are the same everywhere on the sample, and independent of rotation of the sample about the normal axis of the sample.

A small amount of  liquid deposited on this substrate will make an interface with the surrounding gas, which can be described by a second function, $g(x,y)$. The support of $g$  is the set $\mathcal{W} \subset \mathcal{S}$, which denotes the wetted area. Continuity of the liquid surface assures that $g = f$ on the boundary of $\mathcal{W}$, i.e., at the three-phase contact line, where the solid substrate, the liquid, and the gas phase meet. This line will henceforth be denoted by $\partial\mathcal{W}$. 

Information on $g(x,y)$ can be obtained from the total free energy functional of the system, which is given by 
\begin{equation}
F = \int_{\mathcal{W}}\left[(\gamma_{sl}-\gamma_{sg})\sqrt{1+(\nabla f)^2} + \gamma \sqrt{1+(\nabla g)^2} \right] \ {\rm d}x \ {\rm d}y
\label{Eq:TotalFreeEnergy}
\end{equation} 
Minimization of $F$ yields two important properties of $g$. First of all, the mean curvature of the liquid surface, which can be written as \cite{DiffGeom} 
\begin{equation}
H = \frac{1}{2}\nabla\left(\frac{\nabla g}{\sqrt{1+(\nabla g)^2}}\right)
\label{Eq:MeanCurvature}
\end{equation}
assumes the same value everywhere on $\mathcal{W}$. Second, the two surfaces described by $f(x,y)$ and $g(x,y)$ make the same (Young-Dupr\'e) angle $\theta$  everywhere on $\partial\mathcal{W}$. This  reflects the local force balance at the three-phase contact line. 

While surface roughness gives rise to substantial contact angle hysteresis on macroscopic scales, the microscopic contact angle, $\theta$, is known to be well defined on the typical (micrometer to nanometer) scale \cite{SeeBrinkPNAS,TiloPRL}. Nevertheless, we should be aware that even on small scales, equilibration will take time, be it by transport through the gas phase or through an adsorbed layer of molecular thickness \cite{DietrichReview,Seemann2001} (which we disregard in the present study).

The question we shall ask is the following. Given the substrate topography, $f(x,y)$, the equilibrium microscopic contact angle, $\theta$, and the mean curvature of the liquid surface, $H$,  what can we predict on the function $g(x,y)$ and the shape of the wetted area, $\mathcal{W}$? In particular, we shall be interested whether $\mathcal{W}$ forms a percolated set in the plane.

Based on the observation that the amplitude of most natural roughness is much smaller than its dominant lateral length scale, we assume for the present study that 
\begin{equation}
\mid \nabla f \mid \ll 1
\label{Eq:FlatApprox}
\end{equation} 
which allows for substantial simplifications. Expanding then the mean curvature according to eq.~(\ref{Eq:MeanCurvature}), we obtain to first order in $\nabla g$
\begin{equation}
H \approx \frac{1}{2}\Delta g
\label{Eq:MeanCurvAppprox}
\end{equation}
Similarly, 
the contact angle with the substrate yields the boundary condition \begin{equation}
\mid \nabla (g - f)\mid \ \approx \tan\theta
\label{Eq:YoungApprox}
\end{equation}
on $\partial\mathcal{W}$, to first order in $\nabla f$ and $\nabla g$. We can now immediately write down a useful identity concerning these quantities. Green's theorem tells us that  
\begin{equation}
\int_{\partial\mathcal{W}} {\bf n}\nabla (g-f) {\rm d}s= \int_{\mathcal{W}}\Delta (g-f)  {\rm d}^2{\bf x}
\label{Eq:Green}
\end{equation}
where $s$ is the distance along $\partial\mathcal{W}$, ${\bf n}$ its unit normal vector, and ${\bf x} = (x,y)$. Since $g = f$ on $\partial\mathcal{W}$,$\nabla (g-f)$ is everywhere perpendicular to $\partial\mathcal{W}$. Hence  eq.~(\ref{Eq:YoungApprox}) may be written as ${\bf n}\nabla (g-f) \approx \tan\theta$, and we can rewrite eq.~(\ref{Eq:Green}) as
\begin{equation}
L\  \tan\theta + \int_{\mathcal{W}} [2 H -\Delta f] {\rm d}^2{\bf x} = 0 
\label{Eq:Result}
\end{equation}
in which $L$ denotes the length of $\partial\mathcal{W}$. This equation will be the starting point of the discussion to follow.

\section{Gaussian roughness}

If we want to exploit eq.~(\ref{Eq:Result}), we have to refer to a specific roughness function, $f({\bf x})$.  Following the overwhelming majority of the literature on randomly rough surfaces, we will start by considering Gaussian roughness. The height distribution is then 
\begin{equation}
p(f) = \frac{1}{\sqrt{2\pi C_0}}\exp\{-f^2/2C_0\}
\label{Eq:Gauss}
\end{equation}
where $C_0$ is the mean square of $f({\bf x})$.
$f$ can be fully characterized by its lateral correlation function \cite{LongHigg1957,LongHigg1957,Nayak1973,Green1984,Ogilvy1989,Isich1992}, 
\begin{equation}
C(r) = <f({\bf x})f({\bf x} + {\bf r})> 
\label{Eq:Correlation}
\end{equation}
with $r = \mid {\bf r}\mid$. Below we will make use of its polynomial expansion,  
\begin{equation}
C(r) = C_0 - \sum_{n=1}^{\infty} C_n r^n 
\label{Eq:CExpansion}
\end{equation}
where $C_0$ is the mean square amplitude of the roughness. The form of eq.~(\ref{Eq:Correlation}) reflects the isotropy of the roughness \cite{Isich1992}.

\subsection{Distribution functions}

For Gaussian roughness, the joint distributions of $f$ with other stochastic quantities can be obtained in a straightforward manner from multivariate analysis. As it is well known \cite{LongHigg1957,Green1984}, the joint distribution of two quantities $f_1$ and $f_2$ is then given by
\begin{equation}
p(f_1,f_2) = \frac{1}{2\pi\sqrt{Q}}\exp\left[-\frac{1}{2}\sum_{ij}M_{ij} f_i f_j\right]
\label{Eq:Multivariate}
\end{equation}
where $(M_{ij})$ is the inverse of the matrix $(<f_i f_j>)$ and $Q$ is the determinant of that matrix \cite{LongHigg1957}. For the joint probability of $f$ and $\nabla f$, we find \begin{equation}
p(f, \nabla f) = \frac{p(f)}{4\pi C_2}\exp\left[-\frac{(\nabla f)^2}{4C_2}\right]
\label{Eq:JointSlope}
\end{equation}
On the side, this directly yields $<(\nabla f)^2> = 4C_2$, from which we can conclude that roughness topographies fulfilling (\ref{Eq:FlatApprox}) will have $C_2 \le 0.003$. 
 
For the joint probability of $f$ and $\Delta f$, we obtain
\begin{equation}
p(f,\Delta f) = \frac{1}{2\pi\sqrt{Q}}\exp\left[-\frac{64C_4 f^2 - 8 C_2 f \Delta f + C_0 (\Delta f)^2}{2Q}\right]
\label{Eq:JointLaplace}
\end{equation}
with $Q = \mid 64C_0C_4 - 16 C_2^2\mid$.
From these expressions, we can derive a number of useful formulae. For the length of the contour line of $f(x,y)$ at height $f=h$, we find \cite{LongHigg1957b, Nayak1973}
\begin{equation}
L(h) = \int \mid \nabla f \mid \ p(h,\nabla f) \ {\rm d}^2 (\nabla f) = \sqrt{\pi C_2}\ p (h)
\label{Eq:Contour}
\end{equation} 
The fraction of the total sample area which lies within that contour is 
\begin{equation}
W(h) = \int\limits_{-\infty}^h p(f) \  df = \frac{1}{2}\left[1 + {\rm erf} \left(\frac{h}{\sqrt{2C_0}}\right)\right]
\label{Eq:Area}
\end{equation} 
and the total Laplace curvature of $f$ within that area is
\begin{equation}
K(h) = \int\limits_{-\infty}^h \int\limits_{-\infty}^{+\infty} \Delta f \  p(f,\Delta f) \ {\rm d}(\Delta f) \ {\rm d}f  = 4 C_2 p(h)
\label{Eq:Laplace}
\end{equation} 
 
In order to fulfill the boundary condition, eq.~(\ref{Eq:YoungApprox}), the vertical position of the three-phase contact line, which may be symbolically written as $f(\partial\mathcal{W})$, will vary along $\partial\mathcal{W}$ about an average value, $<f(\partial\mathcal{W})>$. The three-phase contact line will thus approximately follow the contour line at $f({\bf x}) = <f(\partial\mathcal{W})>$, with excursions towards both the outside and the inside of $\mathcal{W}$. These will in cases represent detours, sometimes shortcuts with respect to $\partial\mathcal{W}$. As a reasonable approximation, we may thus use $L (<f(\partial\mathcal{W})>)$ for the length of the three-phase contact line. Similarly, we set 
\begin{equation}
\int_{\mathcal{W}} \ {\rm d}{\bf x}  \approx W(h) 
\label{Eq:MCApprox}
\end{equation} 
and 
\begin{equation}
\int_{\mathcal{W}} \Delta f \ {\rm d}{\bf x} \approx K(h)  
\label{Eq:LaplaceApprox}
\end{equation} 
with $h := <f(\partial\mathcal{W})>$. Inserting these expressions in eq.~(\ref{Eq:Result}), we obtain
\begin{equation}
\tan\theta \approx \frac{K(h) - 2 H W(h)}{L(h)}
\label{Eq:Adsorption}
\end{equation}
from which $h$ can be determined. Inserting (\ref{Eq:Contour}) and (\ref{Eq:Laplace}) in eq.~(\ref{Eq:Adsorption}), we arrive at
\begin{equation}
\tan\theta \approx \sqrt{\frac{16 C_2}{\pi}} - \frac{2 H W(h)}{L(h)}
\label{Eq:Adsorption}
\end{equation}

\subsection{The phase diagram}

If the adsorbed material is at liquid-vapor coexistence, the mean curvature of the free liquid surface, $H$, vanishes everywhere on $\mathcal{W}$. In this case, eq.~(\ref{Eq:Adsorption}) is fulfilled only for a certain contact angle, 
\begin{equation}
\theta_c = \arctan(\sqrt{16 C_2 / \pi})
\label{Eq:Thetac}
\end{equation}
Note that $\theta_c$ is independent of $h$. This at first glance puzzling result has its origin in a peculiar property of Gaussian roughness, namely the statistical independence of $\nabla f$ and $f$ \cite{LongHigg1957,Isich1992}. In other words, the probability of finding a certain slope at a given level, $h$, is independent of $h$. A von Neumann boundary condition such as  eq.~(\ref{Eq:YoungApprox}) can thus be fulfilled equally well at all levels of Gaussian roughness. It is therefore not surprising that no particular value of $h$ is singled out here. 

\begin{figure}[h]
\includegraphics[width = 6.5cm]{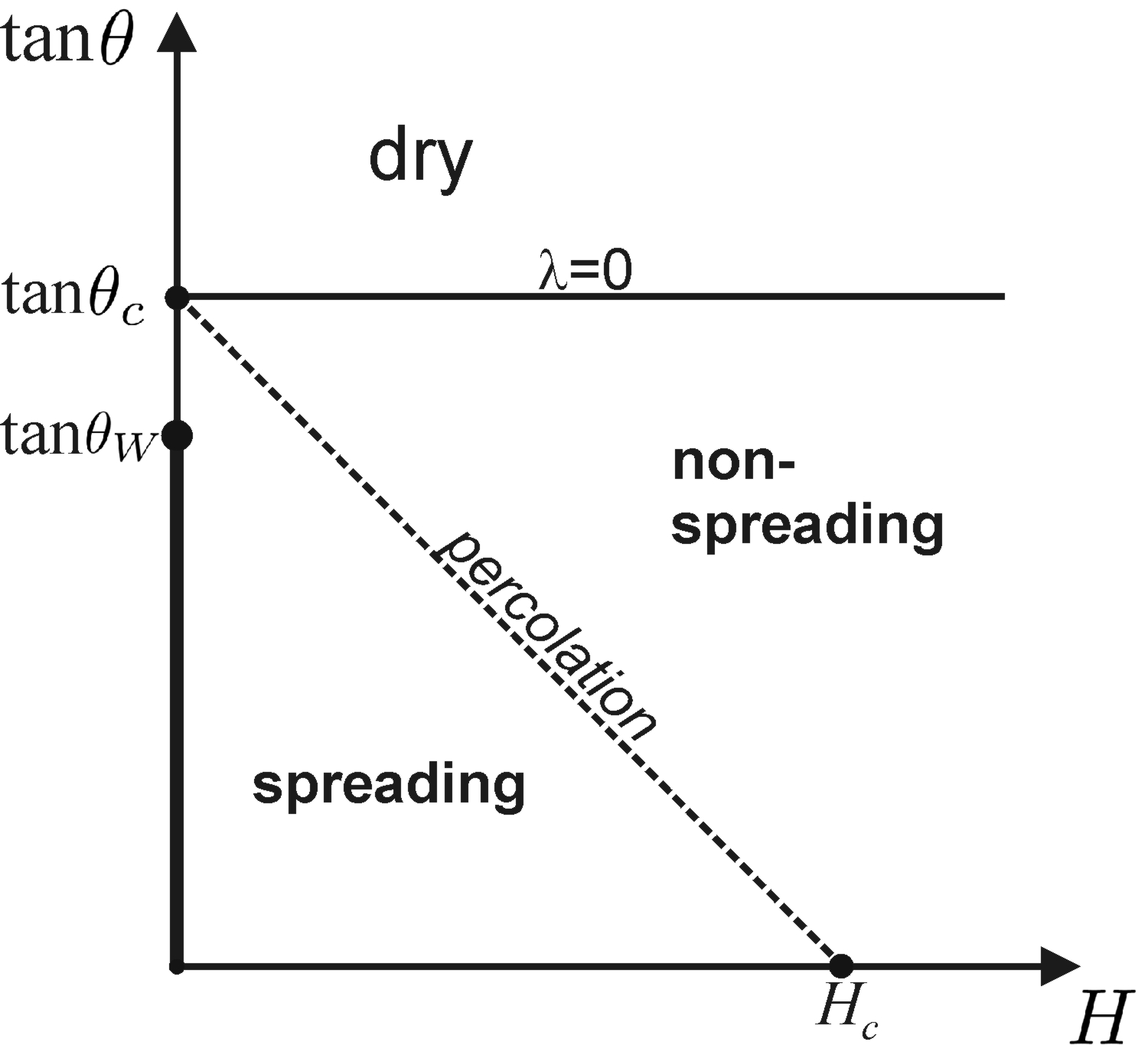}
\caption{The phase diagram of wetting and spreading on a surface with Gaussian roughness. \label{PhaseDiagramGauss} }
\end{figure}

It is instructive to compare $\theta_c$ with $\theta_W$. For Wenzel's parameter ${\rm r}$, we have 
\begin{equation}
{\rm r} = \int_{\mathcal{S}} \sqrt{1+(\nabla f)^2}\ {\rm d}^2{\bf x} \approx 1 + 2C_2 - \mathcal{O}(C_2^2)
\label{Eq:WenzelRatio}
\end{equation}
and therefore $\cos^2\theta_W \approx 1-4C_2$. For $\theta_c$ we obtain, from eq.~(\ref{Eq:Thetac}), $\cos^2\theta_c \approx 1-16C_2/\pi$. Since $16/\pi > 4$, we see that $\theta_c > \theta_W$. In other words, if the liquid does not wet the substrate well enough to fulfill the Wenzel condition, is may nevertheless well intrude the roughness topography and thus form a wetting layer. This is indicated in Fig.~\ref{PhaseDiagramGauss}, which shows the phase diagram of wetting on a surface with Gaussian roughness. States corresponding to liquid/vapour coexistence lie on the vertical axis. Along the bold solid line, which ends at $\theta_W$, the liquid surface can 'detach' from the rough substrate, such that an infinitely thick liquid film may form. For $\theta_W < \theta < \theta_c$, the liquid/vapour interface needs the support of the spikes of the roughness, to which it is attached by virtue of the boundary condition, eq.~(\ref{Eq:YoungApprox}). 

Let us now discuss what happens as we move off coexistence. We first define the parameter 
\begin{equation}
\lambda = \sqrt{16 C_2/\pi} - \tan \theta
\label{Eq:lambda}
\end{equation}
which only depends upon the substrate topography (through $C_2$) and $\theta$. Inserting this into eq.~(\ref{Eq:Adsorption}), we obtain
\begin{equation}
2 H \ W(h) = \lambda \ L(h)
\label{Eq:OffCoex}
\end{equation} 
as an alternative form of (\ref{Eq:Adsorption}).

For Gaussian roughness, it is well known that percolation of the set $\mathcal{P}(h) = \{(x,y)\mid f(x,y) \leq h \}$ takes place at $h=0$ \cite{Isich1992}. From this and eq.~(\ref{Eq:Contour}), (\ref {Eq:Area}), and (\ref{Eq:OffCoex}) we can immediately derive that percolation will take place when 
\begin{equation}
H = \lambda\sqrt{\frac{C_2}{2 C_0}}
\label{Eq:Percolation}
\end{equation}
This is indicated by the dashed straight line in Fig.~\ref{PhaseDiagramGauss}, which for $\theta = 0$ ends at $H_c = \sqrt{8C_2^2/\pi C_0}$. This line cuts through the whole range of contact angles below $\theta_c$. At higher angles, eq.~(\ref{Eq:YoungApprox}) cannot be fulfilled and the substrate remains dry everywhere. 

\subsection{Adsorption isotherms}

We can now calculate the adsorption isotherms of the system, i.e., the amount of liquid adsorbed at pressures below saturation. This is important to discuss, as almost no practical situation corresponds exactly to liquid/gas coexistence. 
Consider, for instance, the substrate to be located at a height $Z$ above a liquid reservoir with which it can exchange material. $H$ is then given by the balance with the hydrostatic pressure and reads
\begin{equation}
H = \frac{\rho g Z}{2\gamma}
\end{equation}
For the distribution of water within a soil or granular pile at height $Z$ above the water table, we find that $H$ grows to about $1/100$ nm as $Z$ increases to $10$ m. Hence the typical curvatures to expect are of the right size for our considerations up to several meters above the water table. 

A more general way to look at this situation is to consider the vapour pressure, which is reduced at finite height above the liquid reservior, as well due to gravity. The curvature is then given by the Kelvin equation, 
\begin{equation}
H = \ln \frac{p_s}{p} \frac{k_B T}{2 \gamma v_m}
\label{Eq:Kelvin}
\end{equation}
where $p$ is the partial pressure of the adsorbed liquid in the sorrounding gas phase, $p_s$ is its saturated vapor pressure, $v_m$ its molecular volume, and $k_B$ is Boltzmann's constant. 

\begin{figure}[h]
\includegraphics[width = 8.5cm]{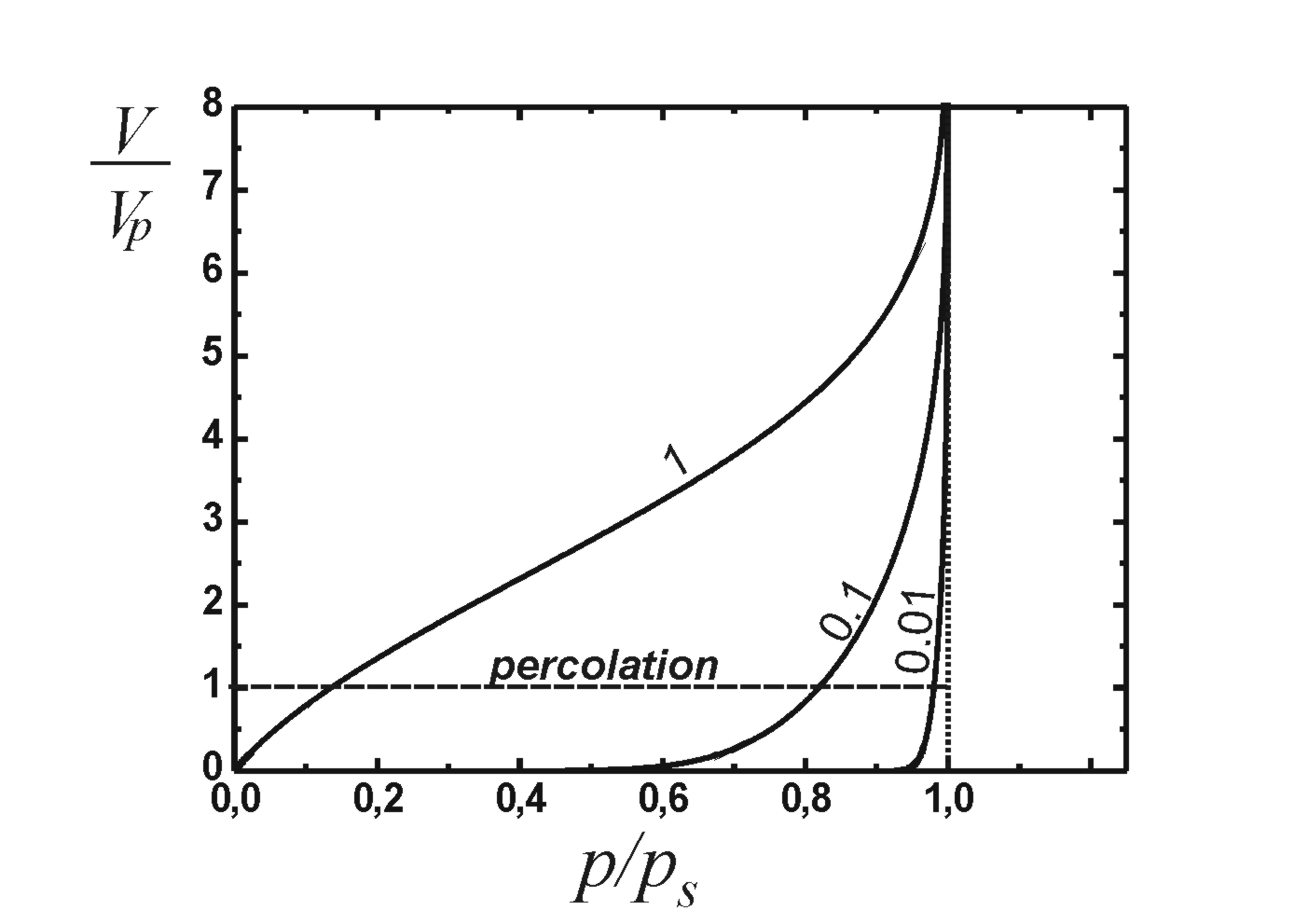}
\caption{Adsorption isotherms on Gaussian roughness. The parameter in the family of curves is $\lambda/\alpha$.\label{Adsorption} }
\end{figure}

From eq.~(\ref{Eq:Kelvin}), with the abbreviation $\alpha = k_B T/\gamma v_m$, we can express the adsorption isotherms in terms of $h$ as
\begin{equation}
\frac{p}{p_s} = \exp \left[-\frac{\lambda L(h)}{\alpha W(h)}\right]
\label{Eq:AdsorptionIsotherm}
\end{equation}
We would, however, like to know not the position of the liquid surface, $h$, but the total volume of adsorbed liquid. The latter can be easily expressed as
\begin{equation}
V = \int\limits_{-\infty}^{h} (h-f) p(f) {\rm d}f = \frac{h}{2} \left[1+{\rm erf}\left(\frac{h}{\sqrt{2C_0}}\right)\right]+\sqrt{\frac{C_0}{2\pi}}e^{-\frac{h^2}{2C_0}}
\label{Eq:AdsorbedVolume}
\end{equation}
The volume at percolation, i.e. at $h = 0$, is $V_p = \sqrt{C_0/2\pi}$. 
Combining eqs.~(\ref{Eq:AdsorptionIsotherm}) and (\ref{Eq:AdsorbedVolume}), we can directly plot the adsorption isotherms, which are displayed in Fig.~\ref{Adsorption} for three different values of $\lambda/\alpha$. If $\lambda = 0$, $V$ remains zero for all $p<p_s$, and jumps to infinity at $p=p_s$. 

It is important here to appreciate that the infinite adsorption one obtains at coexistence has two different meanings for contact angles above or below $\theta_W$. While for $\theta < \theta_W$, the liquid can detach completely from the substrate forming a bulk liquid phase, the liquid surface remains in contact with the rough substrate for $\theta > \theta_W$. The fact that even then the adsorption isotherms tend to infinity at coexistence owes to the infinite support of the Gaussian distribution, eq.~(\ref{Eq:Gauss}). We will see below that this is a peculiarity of Gaussian roughness, and not a generic feature of practically encountered roughness topographies.

\section{Non-Gaussian roughness}

As we have seen so far, it is worthwhile to study non-Gaussian roughness models as well. In fact, it has been shown that many real surfaces are distinctly non-Gaussian \cite{Adler1981,McCool1992,Wu2004,RodValve2010}, such that the freedom in adjusting the correlation function is not sufficient to describe a relevantly large class of surfaces. It seems to be widely believed that the correlation function together with the height distribution of the topography are sufficient to characterize all physically relevant properties of a surface. Most authors even seem to believe that only the first four moments of the height distribution are relevant (including  skewness and kurtosis)  \cite{Adler1981,McCool1992,Wu2004,Bakolas2003}. In what follows, we will introduce a simple roughness model which is general enough to describe roughness profiles with any lateral correlation function and height distribution, but is still accessible to the analysis given above. As a consequence, we will be able to derive, by purely analytic methods, quite general predictions about wetting, adsorption, and liquid percolation on a rough surface, which can be quantitatively applied to experimental data. 

Let $\chi({\bf x})$ be a Gaussian random function, much like $f$ as discussed above, but with dimensionless codomain and unity root mean square. Hence its height distribution is
\begin{equation} 
q(\chi ) = (2\pi )^{-\frac{1}{2}}\exp\left[-\frac{1}{2}\chi^2\right]
\label{Eq:HeightDistrChi}
\end{equation}
and the correlation function,
\begin{equation} 
<\chi({\bf x})\chi({\bf x} + {\bf r})> = 1 - \sum_{m=1}^{\infty} D_m r^m
\label{Eq:CorrelationChi}
\end{equation}
We then set 
\begin{equation}
f({\bf x}) = S(\chi({\bf x}))
\label{Eq:NonGauss}
\end{equation}
where $S$ has the dimension of a length and is a monotone, two times differentiable function. In this case the inverse of $S(\chi)$, $T(f) := S^{-1}(f)$, exists, and we have
\begin{equation}
p(f) = T^{\prime}(f) \ q(T(f))
\label{Eq:ProfileNonGauss}
\end{equation}
where the prime indicates the derivative with respect to the argument. 

\begin{figure}[h]
\includegraphics[width = 8.5cm]{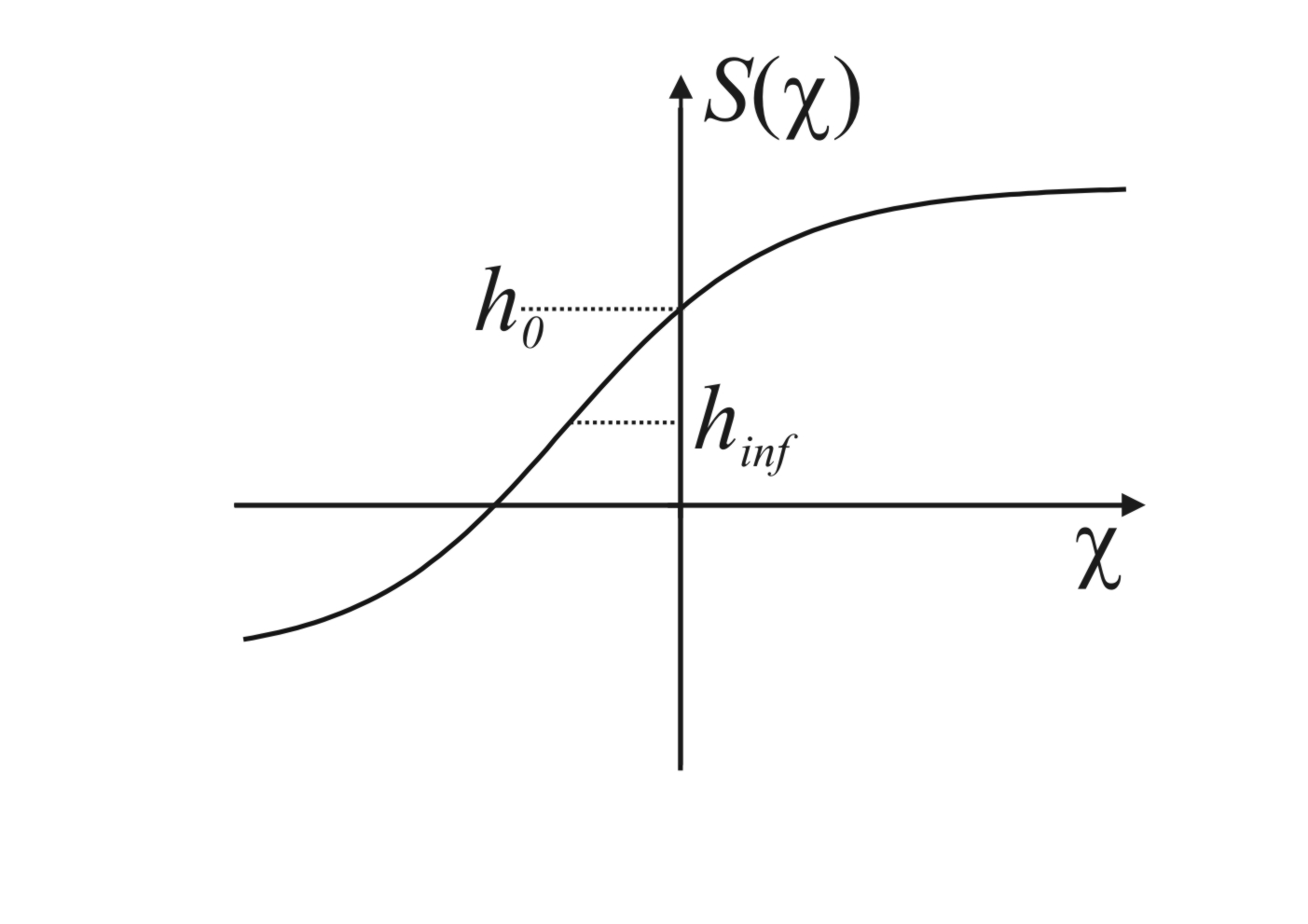}
\caption{Sketch of a typical $S(\chi)$, as would be obtained on a worn surface. While the support of $S$  is the whole $\chi$ axis, the codomain is bound because neither will there be any material outside the original (unworn) surface, nor will there be infinitely deep troughs. $h_{inf}$ indicates the inflection point of $S$. \label{S(chi)} }
\end{figure}

$S(\chi)$ can be directly determined from experimental topography data. If  the distribution $p(f)$ has been  measured, we can derive $T(f)$  by means of the simple formula
\begin{equation}
T(f) = {\rm erf}^{-1}\left[2 \int\limits_{-\infty}^{f}p(z)\ {\rm d}z - 1\right] 
\label{Eq:Transform}
\end{equation}
Note that this allows to represent any height distribution function $p(f)$. The correlation function of $\chi$, and thereby the set of coefficients $D_m$, is obtained from the data as $<T(f({\rm x}))T(f({\rm x + r}))>$. Fig.~\ref{S(chi)} shows a sketch of a typical $S(\chi)$. While the support of $S$  is the whole $\chi$ axis, the codomain is bound, because neither will there be any material outside the original (unworn) surface, nor will there be infinitely deep troughs. 

Since in eq.~(\ref{Eq:NonGauss}) we have done nothing but distorting the assignment of vertical positions to the plane,  contour lines and their enclosed areas will change  in level according to $S(\chi)$, but their topolgical properties, including percolation, will remain unchanged. We can therefore directly write down the contour length with help of the new quantities,
\begin{equation}
L_S(h) = \sqrt{\pi D_2}\ q\left[T(h)\right]
\label{Eq:ContourNonGauss}
\end{equation} 
For the joint probability of $f$ and $\nabla f$, we obtain
\begin{equation}
p(f,\nabla f) = \frac{T^{\prime 3} q(T)}{4\pi D_2}\exp\left[- \frac{T^{\prime 2} (\nabla f)^2}{4 D_2}\right ]
\label{Eq:fsigmaNonGauss}
\end{equation} 
where the prime now denotes the derivative with respect to the argument. The Laplace curvature is given by $\Delta f = S^{\prime \prime} (\nabla \chi)^2 + S^{\prime} \Delta \chi$, which leads to intimidatingly clumsy expressions when inserted into multivariate analysis. We therefore consider here the important case when $ S^{\prime \prime}$ is small, such that only the second term in $\Delta f$ contributes. This is the case if
\begin{equation}
\mid\frac{ S^{\prime \prime}}{ S^{\prime  }}\mid \ll \frac{\sqrt{<(\Delta f)^2>}}{<(\nabla f)^2>} = \frac{2\sqrt{D_4}}{D_2}
\label{Eq:SppSmall}
\end{equation}
We then have 
\begin{equation}
p(f,\Delta f) = \frac{T^{\prime 2}}{2\pi \sqrt{R}}\exp\left[-\frac{64 D_4 T^2 -8 D_2 T T^{\prime}\Delta f + (T^{\prime}\Delta f)^2}{2R} \right] 
\label{Eq:fkappaNonGauss}
\end{equation} 
with $R = \mid 64 D_0 D_4 - 16 D_2^2\mid $. Now we are in shape to express the Laplace curvature inside the wetted area. In complete analogy to the derivation above, we find
\begin{equation}
K_S(h) = \frac{4D_2}{T^{\prime}(h)}q\left[T(h)\right]
\label{Eq:KNonGauss}
\end{equation} 
At coexistence, we have again 
\begin{equation}
L_S(h) \tan \theta = K_S(h)
\label{Eq:Balance}
\end{equation}
In analogy to the above discussion, we define the parameter 
\begin{equation}
\Lambda := \frac{4}{T^{\prime}(h)}\sqrt{\frac{D_2}{\pi}}-\tan\theta
\label{Eq:LambdaNonGauss}
\end{equation} 
which this time does depend on $h$, as is sketched in Fig.~\ref{Lambda(h)}. The maximum of the curve lies at $h_{inf}$, which is the inflection point of $S(\chi)$ (cf. Fig.~\ref{S(chi)}). 

\subsection{The phase diagram}

The film thickness at coexistence can be derived from the zeros of $\Lambda$, of which there are either two or none. In the latter case, the contact angle (and thereby $\tan\theta$) is too large for forming a liquid surface between the spikes and troughs which complies with the boundary condition, eq.~(\ref{Eq:YoungApprox}). If, however, $\Lambda$ intersects the $h$-axis, the slope of the zeros decides upon their stability. This can be seen by appreciating that $\Lambda$ may be interpreted as a deviation from the force balance, eq.~(\ref{Eq:Balance}), as required by eq.~(\ref{Eq:YoungApprox}). For the left zero, which is marked by an open circle in the figure, a displacement of the three-phase contact line would give rise to an imbalance of wetting forces driving it further away from the zero. The opposite is true for the right zero, marked by the closed circle. The latter is therefore stable and thus corresponds to the adsorbed film thickness which will develop. The film will be percolated if this zero lies to the right of $h_0 = S(0)$, which corresponds to the mid-plane of $\chi({\bf x})$ (cf. Fig.~\ref{S(chi)}). 

\begin{figure}[h]
\includegraphics[width = 8.5cm]{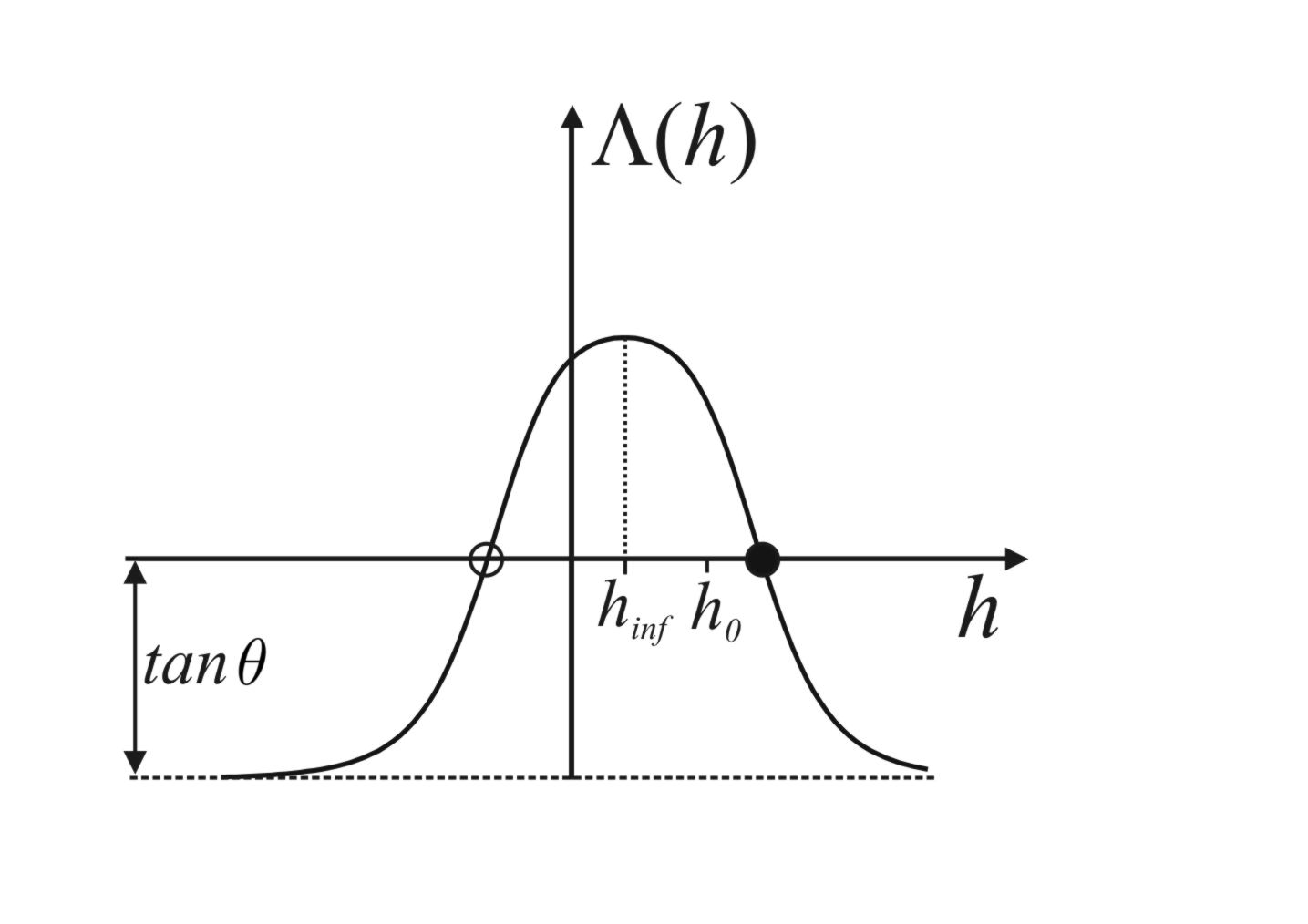}
\caption{Sketch of $\Lambda(h)$ corresponding to $S(\chi)$ as in Fig.~\ref{S(chi)}. The film thickness at coexistence corresponds to the zero with negative slope (closed circle). \label{Lambda(h)} }
\end{figure}

\begin{figure}[h]
\includegraphics[width = 8.5cm]{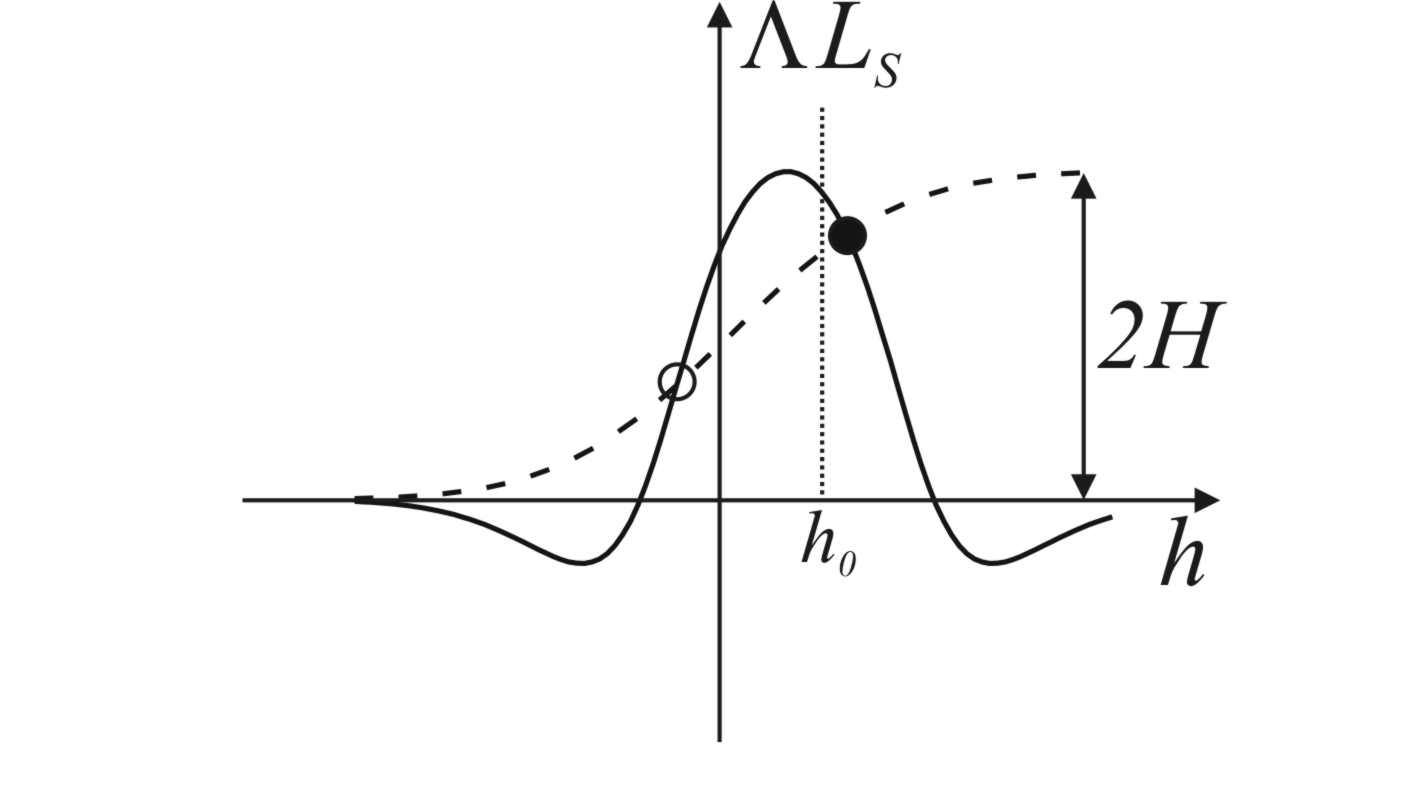}
\caption{Graphic construction for solving eq.~(\ref{Eq:OffCoexNonGauss}).  The dashed line represents the l.h.s. of   eq.~(\ref{Eq:OffCoexNonGauss}). \label{LambdaL} }
\end{figure}

\begin{figure}[h]
\includegraphics[width = 8.5cm]{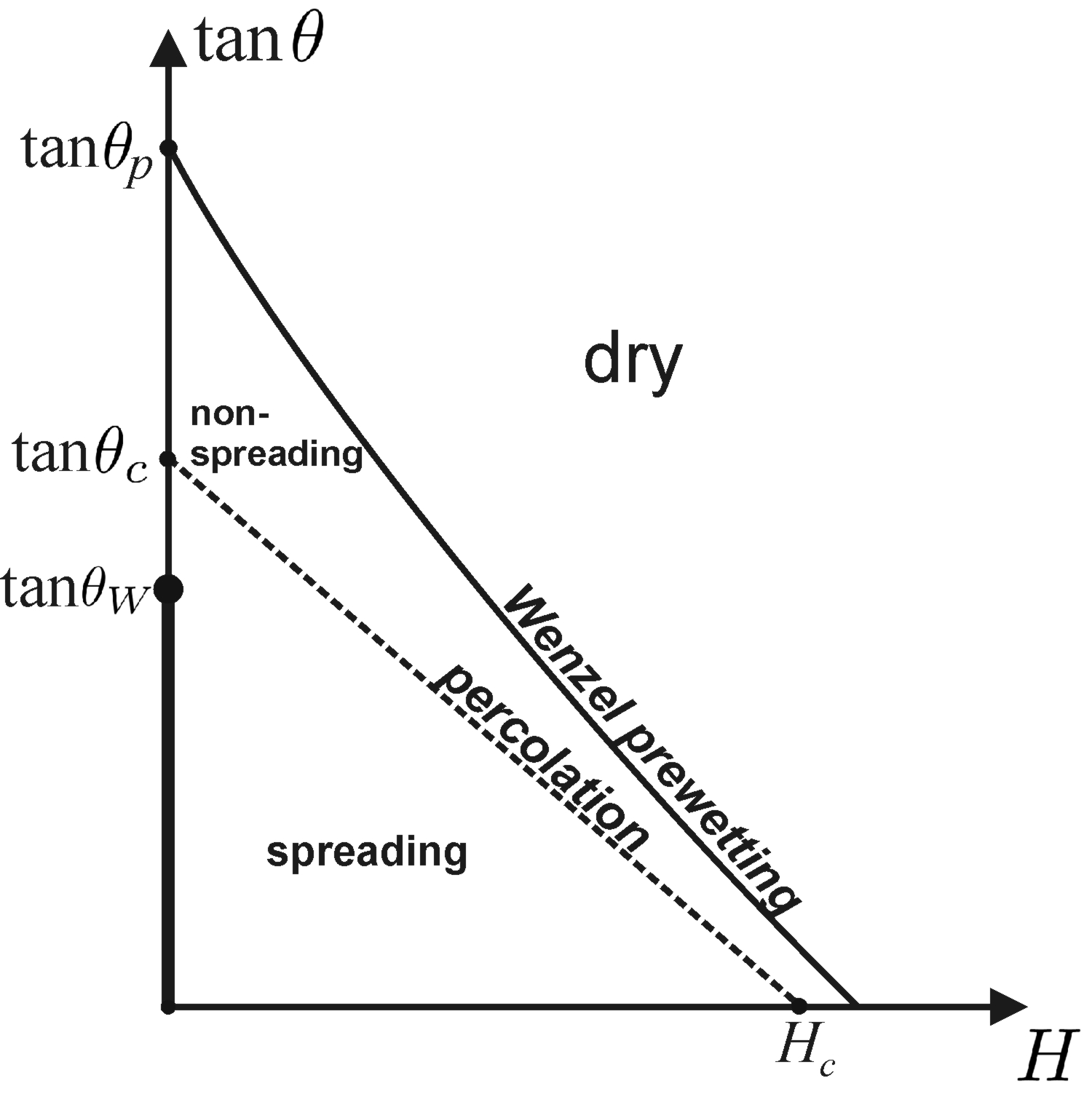}
\caption{The wetting phase diagram for non-Gaussian roughness. The most prominent qualitative difference with respect to Gaussian roughness is the existence of a 'Wenzel prewetting line', at which the adsorbed film thickness jumps discontinuously from zero (or, in fact, a molecularly thin adsorption layer) to a finite value.} 
\label{PhaseDiagramNonGauss}
\end{figure}

If we now again  consider the system off coexistence, we have 
\begin{equation}
2 H\ W_S(h) = \Lambda (h) \ L_S(h)
\label{Eq:OffCoexNonGauss}
\end{equation} 
as the condition for $h$, where 
\begin{equation}
W_S(h) = \frac{1}{2}\left[1+{\rm erf} \left(\frac{T(h)}{\sqrt{2}}\right)\right]
\label{Eq:WS(h)}
\end{equation}
A graphical solution of eq.~(\ref{Eq:OffCoexNonGauss}) is sketched in Fig.~\ref{LambdaL}. Again, the closed circle indicates the stable solution. The liquid film will be percolated if this point lies to the right of the dashed line at $h_0$, but form isolated patches otherwise. From eq.~(\ref{Eq:OffCoexNonGauss}), we see that percolation occurs if \begin{equation}
H = \sqrt{\frac{D_2}{2}}\Lambda (h_0)
\label{Eq:PercolationNonGauss}
\end{equation}
which represents again a straight line in the phase diagram as depicted in Fig.~\ref{PhaseDiagramNonGauss}. For Gaussian reoughness, we have $S(\chi) = \sqrt{C_0}\chi$, $D_m = C_m/C_0$, and $T^{\prime} = 1/\sqrt{C_0}$. It is readily checked that this leads again to eq.~(\ref{Eq:Percolation}) instead of (\ref{Eq:PercolationNonGaus}), and (\ref{Eq:lambda}) instead of (\ref{Eq:LambdaNonGauss}).

As in the case of Gaussian roughness, $\theta_W$ generically lies below $\theta_c$. This can be seen from calculating 
\begin{equation}
(\cos\theta_W)^{-2} = {\rm r} -1 \approx 4 D_2 \int q(\chi)S^{\prime 2}(\chi){\rm d}\chi
\label{Eq:WenzelNonGauss}
\end{equation}
which follows from (\ref{Eq:fsigmaNonGauss}). On the other hand, 
\begin{equation}
(\cos\theta_c)^{-2} = \frac{16 D_2}{\pi} S^{\prime 2}(0)
\label{Eq:ThetacNonGauss}
\end{equation}
Since, again, $16/\pi > 4$, it is clear that whenever ${\rm argmax}(S^{\prime}) = T(h_{inf})$ lies close to $0$ (which it typically will), we have $\theta_c > \theta_W$ as for the Gaussian case (cf. Fig.~\ref{PhaseDiagramNonGauss}). 

Inspection of Fig.~\ref{LambdaL} shows that the two points of intersection, which are marked by the closed and open circles, will merge when the dashed and solid curves touch each other in a single point. This occurs at a certain curvature $H_p(\theta)$ of the liquid surface. For $H > H_p$, there is no liquid adsorbed, and the substrate is dry. As $H_p$ is reached, the average position of the liquid surface, $h$, jumps discontinuously to the value given by the point of contact of the two curves. As $H$ is further reduced,  $h$ increases until at coexistence it reaches a value corresponding to the right zero of $\Lambda L_S$. Because of the phenomenological similarity of the jump in adsorbed film thickness to the prewetting transition encountered in standard wetting scenarios on flat substrates \cite{DietrichReview}, we hereby propose to term this transition 'Wenzel prewetting'. When the microscopic contact angle is varied, a 'Wenzel prewetting line' results, which is shown in Fig.~\ref{PhaseDiagramNonGauss} as the solid curve. As in the usual prewetting scenario, this line ends in a critical end point, when the solid and dashed curves in Fig.~\ref{LambdaL} intersect in only a single point. It is readily appreciated from the construction sketched in Figs.~\ref{Lambda(h)} and \ref{LambdaL}, however, that this can occur only for $\theta \le 0$, and thus outside the physically accessible  parameter range. In principle, the Wenzel prewetting line may intersect the percolation line. The latter then follows the prewetting line down to $\theta = 0$. 

\begin{figure}[h]
\includegraphics[width = 8.5cm]{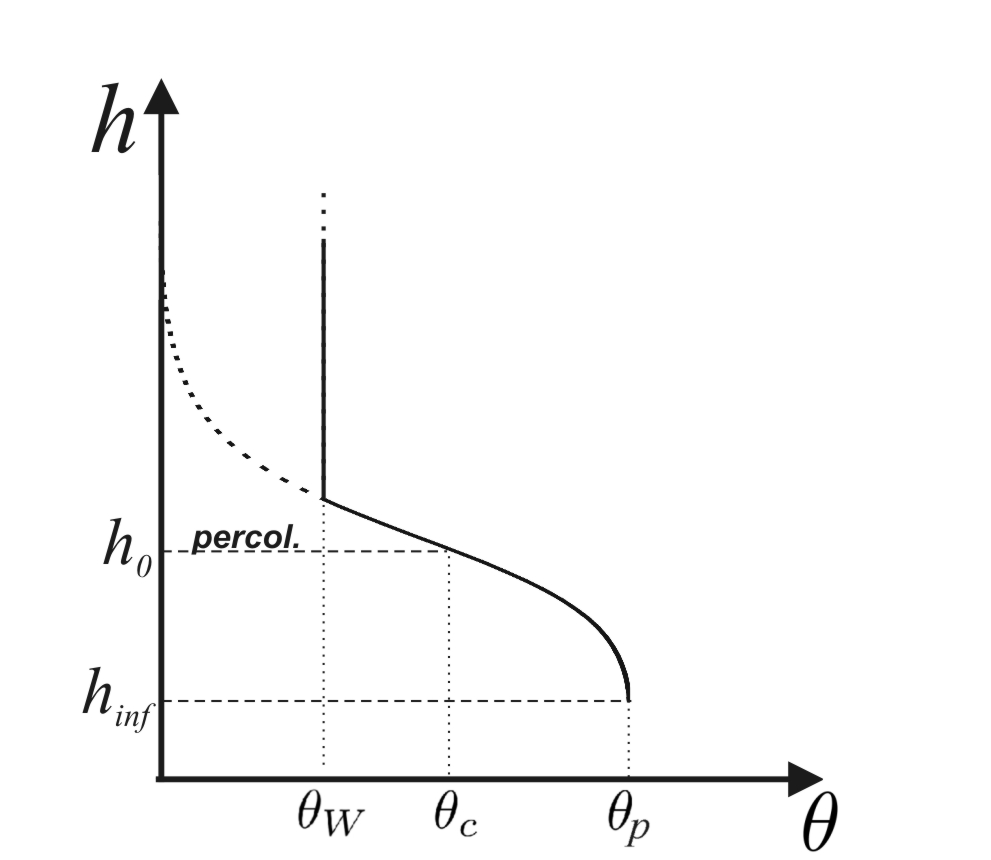}
\caption{ \label{h(theta)} }
\end{figure}

Let us discuss how the position of the liquid surface varies along liquid/vapour coexistence as $\theta$ is gradually decreased from above $\theta_p$. This can be directly read off Fig.~\ref{Lambda(h)}, by inverting $\Lambda(h)$ for $h > h_{inf}$, and is sketched in Fig.~\ref{h(theta)}. The liquid film first appears through a discontinuous jump as $\theta$ crosses the Wenzel prewetting line. As $\theta < \theta_W$, the liquid surface configuration which is bound to the surface topography through eq.~(\ref{Eq:YoungApprox}) becomes metastablee (dashed curve), and the global minimum of the total free energy corresponds to the 'detached' liquid surface, or bulk liquid adsorption (bold line in Fig.~\ref{h(theta)}). 

It is tempting to try to calculate the macroscopic contact angle, $\theta_{macro}$, along the coexistence line for $\theta_W < \theta < \theta_p$. In that range, the fraction $W_S(h)$ of the sample is covered with liquid, while the remaining fraction, $1-W_S(h)$, still exposes the uncovered rough substrate. The liquid surface energy of the areas covered with liquid is $W_S(h) G(h)$, where $G(h) = <\sqrt{1+(\nabla g)^2}>_{\mathcal{W}}$ is the total liquid surface area over $\mathcal{W}$. With the help of (\ref{Eq:WenzelNonGauss}) we readily obtain
\begin{equation}
\cos \theta_{macro} = G(h) - \int\limits_{T(h)}^{\infty} q(\chi) \left[G(h)-4D_2 \cos \theta \ S^{\prime 2}(\chi)  \right]{\rm d}\chi
\label{Eq:ContactAngleNonGauss}
\end{equation}
Unfortunately, there is no straightforward way to calculate $G(h)$. We thus content ourselves here with an upper bound for $\theta_{macro}$, which is obtained by setting $G = 1$ in the above expression. Qualitatively, we can nevertheless conclude that since the jump at the prewetting line will directly enter in the lower boundary of the integral, it is clear that this jump will as well be visible in $\theta_{macro}(\theta)$. This is in contrast to, e.g., first order wetting, where a jump in film thickness is accompanied by a continuous variation in the contact angle \cite{DietrichReview}. We mention again that $\theta_{macro}$ may be subject to significant contact angle hysteresis \cite{Joanny1984} unless long equilibration times are taken into account. 

\subsection{Adsorption isotherms} 

\begin{figure}[h]
\includegraphics[width = 7.5cm]{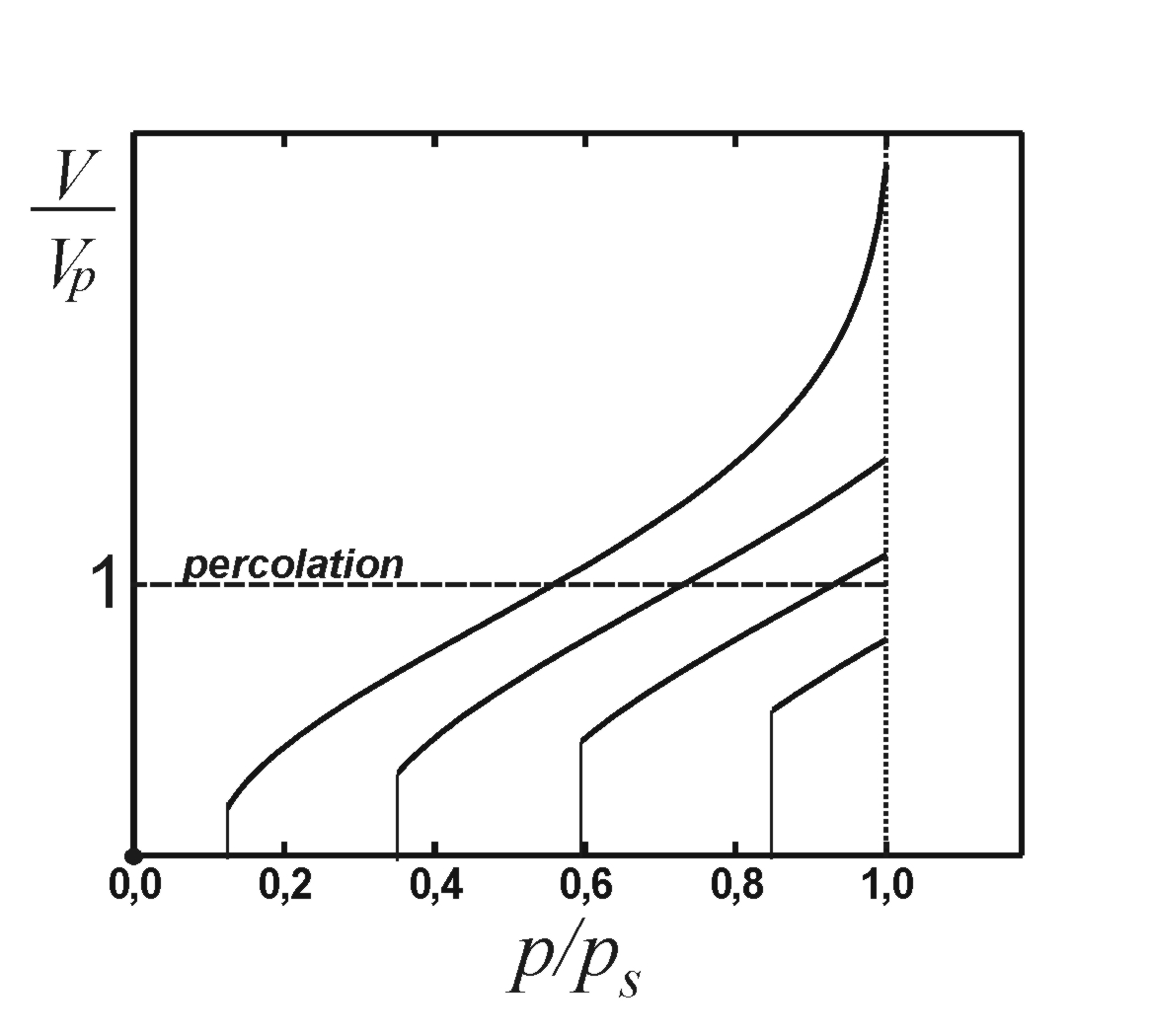}
\caption{Sketch of a family of adsorption isotherms on non-Gaussian roughness. The jump occurs at what we shall term the 'Wenzel prewetting line'. The curves correspond, from bottom to top, to $\theta_c < \theta < \theta_p$, $\theta_W < \theta < \theta_c$, $0 < \theta < \theta_W$, and $\theta = 0$. The latter curve ends at finite adsorbed volume, but with infinite slope. \label{AdsorptionNonGauss}}
\end{figure}

It is finally instructive to discuss the qualitative shape of the adsorption isotherms in this scenario, which are sketched in Fig.~\ref{AdsorptionNonGauss}. The curves, which are meant to correspond to different values of $\theta$, follow what one would expect for the characteristic shown in Fig.~\ref{S(chi)}. The jump from zero film thickness to a finite value is generic and occurs for all contact angles. As $\theta$ is increased, the height of the jump increases slightly, following the curvature of the maximum of $\Lambda(h)$. At the same time, the maximum film thickness (reached at coexistence) decreases, until it finally comes below the percolation threshold when $\theta > \theta_c$. When $\theta > \theta_p$, $\Lambda(h)$ has no zero anymore, and the substrate remains dry up to coexistence. 

A few more words concerning the shape of the function $\Lambda (h)$ are in order. If $f({\bf x})$ were Gaussian, $S(\chi)$ would be just of the form $\sqrt{C_0} \chi$. In that case, $\Lambda (h)$ would in Fig.~\ref{Lambda(h)} be represented by a horizontal line above the abscissa. The solid curve in Fig.~\ref{LambdaL} would then be a Gaussian, and the adsorption isotherms would of course be the same as in Fig.~\ref{Adsorption}. However, the fact that any real roughness is bounded, as there are neither infinitely high spikes nor infinitely deep troughs, entails the boundedness of the codomain of $S(\chi)$, in contrast to the infinite codomain of $\chi$. As an immediate consequence, the derivative of $T(h)$ must finally diverge at the boundaries of its (finite!) support, which necessarily leads to $\Lambda$ bending down onto the dashed line below the abscissa in Fig.~\ref{Lambda(h)}. This leads not only naturally to a finite $h$ at coexistence ($H=0$, cf. Fig.~\ref{LambdaL}), but also to the Wenzel prewetting jump in $h(H)$ farther away from coexistence, when the dashed curve in Fig.~\ref{LambdaL} just touches the solid curve. This reveals that the shape of the adsorption isotherms we derived for Gaussian roughness above is qualitatively different from what should be expected for real roughness. In fact, it misses the whole prewetting scenario, which turned out above to be a generic feature. Once again, Gaussian roughness reveals itself as a special case, which is mathematically convenient but may be misleading when it comes to making quantitative predictions.

\section{Conclusions}

In conclusion, an analytic theory was presented which allows to calculate the wetting phase diagram, adsorption isotherms, and percolation threshold of the adsorbed liquid film for isotropic, randomly rough substrates with arbitrary lateral correlation function and height distribution. The results are found to depend only upon a few key parameters, which can be clearly identified and derived from experimental sample profile data. We have seen that wetting 'physical' roughness displays a number of features which are not present for exactly Gaussian roughness, such as a prewetting transition occurring well before the Wenzel angle is reached. This could be traced down to subtle properties of Gaussian random functions, which reveal their unphysical nature only at second glance. 

Since for most quantities of interest we could come up with closed analytic expressions, these results may be particularly useful for practical applications. The range of validity of the present theory extends from a few nanomeres up to roughly a millimeter, well below the capillary length of the liquid. The field of such applications is vast, including almost all situations in which a liquid comes into contact with a naturally rough surface. In particular, ramifications of wetting phase transitions, which inherently involve small contact angles, are to be expected and can now be accounted for in closed form.
 
Given the potential relevance of the results presented here, it will be worthwhile to work on relaxing the five major approximations we have used: 
\begin{enumerate}
\item{We have assumed the substrate to be chemically homogeneous.}
\item{We have assumed shallow profiles; eqs.~(\ref{Eq:FlatApprox}) and (\ref{Eq:MeanCurvAppprox}).}
\item{We have neglected higher correlations in the shape of $\partial\mathcal{W}$; eqs.~(\ref{Eq:MCApprox}), (\ref{Eq:LaplaceApprox}), and the paragraph before.}
\item{We have assumed the curvature of the roughness characteristic $S(\chi)$ to be small; eq.~(\ref{Eq:SppSmall}).}
\item{We have assumed isotropy of the roughness; eq.~(\ref{Eq:CExpansion}).}
\end{enumerate} 
The last two are probably the simplest to tackle, while the first one appears as the most difficult to overcome. It should finally be noted that in experiments, one has to reckon with quite long relaxation times for the measured quantities, because at all levels of the roughness there are saddle points \cite{LongHigg1957,Nayak1973}, which act as effective pinning sites for $h$. Equilibration will nevertheless proceed  within manageable time, either via the vapour phase or via the molecularly thin adsorbed film \cite{Seemann2001}.

Inspiring discussions with Daniel Tartakovsky, Siegfried Dietrich, Martin Brinkmann, J\"urgen Vollmer, Sabine Klapp, and Daniela Fliegner are gratefully acknowledged. The author furthermore acknowledges generous support form BP International.

\end{document}